\documentclass[
12pt,
preprint,preprintnumbers,nofootinbib,
groupedaddress,superscriptaddress,amsmath,amssymb]{revtex4}
\usepackage{graphicx}
\usepackage{dcolumn}
\usepackage{bm}
\usepackage{amssymb}
\usepackage{amsmath}
\usepackage{epsfig}    
\usepackage{color}
\usepackage{hhline}

\def\be{\begin{equation}}
\def\ee{\end{equation}}
\newcommand{\bea}{\begin{eqnarray}}
\newcommand{\eea}{\end{eqnarray}}
\newcommand{\nn}{\nonumber}

\numberwithin{equation}{section}

\begin{document}

{\begin{flushright}{EPHOU-18-004, KIAS-P18042,  APCTP Pre2018 - 011}
\end{flushright}}

\title{Predictive neutrino mass textures with origin of flavor symmetries}

\author{Tatsuo Kobayashi}
\email{kobayashi@particle.sci.hokudai.ac.jp}
\affiliation{Department of Physics, Hokkaido University, Sapporo 060-0810, Japan}

\author{Takaaki Nomura}
\email{nomura@kias.re.kr}
\affiliation{School of Physics, KIAS, Seoul 02455, Korea}

\author{Hiroshi Okada}
\email{okada.hiroshi@apctp.org}
\affiliation{Asia Pacific Center for Theoretical Physics, Pohang, Geyoengbuk 790-784, Republic of Korea}

\date{\today}

\begin{abstract}
We investigate origins of predictive one-zero neutrino mass textures in a systematic way. Here we search Abelian continuous(discrete) global symmetries, and non-Abelian discrete symmetries, and show how to realize these neutrino masses. Then we propose a concrete model involving a dark matter candidate and an extra gauge boson, and show their phenomenologies.
\end{abstract}
\maketitle
\newpage

\section{Introduction}
One of the most important issues in particle physics is to solve the mystery of 
the flavor structure of quarks and leptons such as the generation number, mass hierarchy, 
mixing angles, and CP phases.
Indeed, a huge number of studies have been done through various approaches.
The texture Ansatz is one of the interesting approaches.
(See for a review, e.g.  \cite{Fritzsch:1999ee}.)
By assuming a certain mass texture, one can derive several predictions 
among masses and mixing angles as well as CP phases.

The experimental data on the neutrino sector has become more precise by neutrino oscillation experiments, 
although there remain still unknown aspects on the neutrino sector, e.g. 
the absolute values of neutrino masses and the question whether neutrinos are Majorana or Dirac fermions.
Thus, it would be interesting to apply the texture Ansatz to the lepton sector.
Actually, a lot of authors have historically been analyzing neutrino mass textures in various setups.
For example, it is known that only seven neutrino mass patterns (two-zero textures) can predict neutrino oscillation data without conflict of current neutrino oscillation data~\cite{Forero:2014bxa} in the case that neutrinos are Majorana fermions with rank three mass matrix~\cite{Fritzsch:2011qv}.

Recently type-I seesaw models with maximally restricted texture zeros have been systematically 
classified and  analyzed numerically in Refs.~\cite{Barreiros:2018ndn, Rink:2016vvl},
where {\it charged-lepton mass matrix is assumed to be diagonal and only two families of right-handed neutrinos have 
Dirac mass terms with three active neutrinos.}
Then, the active neutrino mass matrix has one texture zero, and obviously one of active neutrinos is massless.
Such patterns with one texture zero lead to several 
interesting predictions among neutrino masses and mixing angles.
Indeed, such predictions for the normal hierarchy are not compatible with the experimental data.
Also, some of patterns with one texture zero for the inverted hierarchy are already ruled out by 
experiments, while others are compatible.

Although the texture Ansatz is quite interesting as mentioned above, 
however, it is unclear why such a pattern of mass matrix is realized.
Our purpose is to explore origins of the neutrino mass textures obtained in Refs.~\cite{Barreiros:2018ndn, Rink:2016vvl}.
In this paper, in order to realize those textures, 
we apply flavor symmetries such as global $U(1)$ symmetries, discrete Abelian symmetries $Z_N$, 
and non-Abelian discrete symmetries.
The flavor symmetries would provide a hint to explore underlying theory 
beyond the standard model (SM).

Indeed non-Abelian discrete flavor symmetries have been studied by a lot of authors 
 in order to realize the lepton masses and mixing angles as well as CP phases.
(See for review Refs.~ \cite{Altarelli:2010gt,Ishimori:2010au,King:2013eh}.)
Furthermore,  
it has been shown that some non-Abelian discrete flavor symmetries appear in 
superstring theory with certain compactifications.
Heterotic string theory on toroidal $Z_N$ orbifolds can lead 
non-Abelian flavor symmetries, e.g. $D_4$, and $\Delta (54)$ \cite{Kobayashi:2006wq}.
(See also \cite{Kobayashi:2004ya,Ko:2007dz}.)\footnote{
In Ref.\cite{Beye:2014nxa}, a relation between gauge symmetries and non-Abelian flavor symmetries is discussed.}
Similarly flavor symmetries can be realized in 
 magnetized D-brane models and intersecting D-brane models within the framework of type II superstring theory 
\cite{Abe:2009vi,BerasaluceGonzalez:2012vb}.
In addition, these flavor symmetries may be subgroups of the modular symmetry in superstring theory 
\cite{Kobayashi:2018rad}.
Thus, flavor symmetry would make a bridge between the neutrino physics 
and underlying high energy physics.

The minimal non-Abelian discrete symmetry is $S_3$ and the next one is $D_4$.
Thus, in this paper we consider these $S_3$ and $D_4$ flavor symmetries as well as global $U(1)$ symmetry to realize 
the neutrino mass textures obtained in Refs.~\cite{Barreiros:2018ndn, Rink:2016vvl}.
We will show that one can realize the desired textures by the $D_4$ flavor symmetry and $U(1)$ symmetry, but not 
by the $S_3$ flavor symmetries.
Also it will be found that the $U(1)$ models need more Higgs fields than the $D_4$ flavor models.
Then, we study  the $D_4$ flavor model by using a concrete model.\footnote{See for models with the $D_4$ flavor 
symmetry, e.g.  Refs.~\cite{Grimus:2003kq,Grimus:2004rj,Ishimori:2008gp,Adulpravitchai:2008yp,
Hagedorn:2010mq,Meloni:2011cc}.}

This paper is organized as follows.
In Sec.~\ref{sec:review},   we give a brief review on the neutrino mass textures classified in Refs.~\cite{Barreiros:2018ndn, Rink:2016vvl}.
In Sec.~\ref{sec:realization}, we study their realization by applying Abelian symmetries  and non-Abelian discrete symmetries. 
In Sec.~\ref{sec:model}, we propose a concrete model, in which we formulate the  boson sector, fermion sector, 
and dark matter sector (DM),
and analyze collider physics based on an additional gauge symmetry. Then
we discuss the DM candidate.  
Finally we conclude and discuss in Sec.~\ref{sec:conclusion}.

\section{Neutrino mass textures}
\label{sec:review}

In this section, we review the neutrino mass textures 
obtained in Refs.~\cite{Barreiros:2018ndn, Rink:2016vvl}.
We consider the flavor basis, where charged lepton mass matrix is diagonal.
Also we study the models, that only two families of right-handed neutrinos have 
Dirac mass terms with three families of left-handed neutrinos.

Active neutrino mass matrix is supposed to be induced from canonical mechanism; $m_\nu\approx m_D M_N^{-1} m_D^T$
after the spontaneously electroweak symmetry breaking. Here $m_D$ is $(3 \times 2)$ Dirac mass matrix and $M_N$ is $(2 \times 2)$ Majorana mass matrix
that come from the following Lagrangian; $y_{D_{ij}} \bar L_{L_i} \tilde H_{SM} N_{R_j} + M_{N_{ij}} \bar N^c_{R_i} N_{R_j}$, where
$\tilde  H_{SM} \equiv (i \sigma_2)  H_{SM}^*$ with the second Pauli matrix $\sigma_2$,  $ H_{SM}$ is the SM Higgs, and $N_R$ are right-handed neutrinos.
Then the neutrino mass matrix can be diagonalized by an unitary matrix $U_{\rm PMNS}$ as
\begin{align}
& U_{\rm PMNS}^T m_\nu U_{\rm PMNS}={\rm diag}(m_1,m_2,m_3),\label{eq:neut-diag}\\
U_{\rm PMNS}&=
\left[\begin{array}{ccc}
c_{12}c_{13} & s_{12}c_{13} & s_{13}e^{-i\delta} \\ 
-s_{12}c_{23}-c_{12}s_{23}s_{13}e^{i\delta} &c_{12}c_{23}-s_{12}s_{23}s_{13}e^{i\delta} & s_{23}c_{13} \\ 
s_{12}s_{23}-c_{12}c_{23}s_{13}e^{i\delta} & -c_{12}s_{23}-s_{12}c_{23}s_{13}e^{i\delta} & c_{23}c_{13} \\ 
\end{array}\right]
\left[\begin{array}{ccc}
1& 0 & 0 \\ 
0 & e^{i\alpha/2} & 0\\ 
0 & 0 & 1 \\ 
\end{array}\right],
\end{align}
where $m_{1,2,3}$ are neutrino mass eigenvalues, which are positive real, $c(s)_{12,13,23}\equiv \cos(\sin)\theta_{12,13,23}$ are the three mixing angles, $\delta$ is the Dirac CP phase, and $\alpha$ is the Majorana phase.
Note here that there exists only one Majorana phase due to reduced $M_N$.

For the Dirac mass matrix $m_D$, the maximally allowed number of texture zeros is  one or two.
Then, such matrices $m_D$ are classified as 
~\cite{Barreiros:2018ndn}
\begin{align}
&T_1: \left[\begin{array}{cc}0  &  \times \\ \times & 0  \\ \times &\times   \\ \end{array}\right],\
T_2: \left[\begin{array}{cc}0  &  \times \\ \times & \times  \\ \times &0   \\ \end{array}\right],
T_3: \left[\begin{array}{cc}\times  &  \times \\ 0 & \times  \\ \times &0   \\ \end{array}\right],
U_1: \left[\begin{array}{cc}\times  &  \times \\ 0 & \times  \\ \times &\times   \\ \end{array}\right],
U_2: \left[\begin{array}{cc}\times  &  \times \\ \times & \times  \\ 0 & \times   \\ \end{array}\right],
\nn\\
&T_4: \left[\begin{array}{cc}\times  &  0 \\ 0 & \times  \\ \times &\times   \\ \end{array}\right],\
T_5: \left[\begin{array}{cc}\times  &  0 \\ \times & \times  \\ 0 &\times   \\ \end{array}\right],
T_6: \left[\begin{array}{cc}\times  &  \times \\ \times & 0  \\ 0 &\times   \\ \end{array}\right],
U_3: \left[\begin{array}{cc}\times &  \times \\ \times & 0  \\ \times &\times   \\ \end{array}\right],
U_4: \left[\begin{array}{cc}\times  &  \times \\ \times & \times  \\ \times & 0   \\ \end{array}\right].
\end{align}
For the right-handed neutrino Majorana mass matrix $M_N$, the maximally allowed number of 
texture zeros is one or two.
Then, such matrices $M_N$ are classified as ~\cite{Barreiros:2018ndn}
\begin{align}
&R_1: \left[\begin{array}{cc} \times &  0 \\ 0 & \times   \\ \end{array}\right],\
R_2: \left[\begin{array}{cc} 0 &  \times \\ \times & \times   \\ \end{array}\right],\
R_3: \left[\begin{array}{cc}\times  &  \times \\ \times & 0   \\ \end{array}\right]. \
S: \left[\begin{array}{cc}0  &  \times \\ \times & 0   \\ \end{array}\right].
\end{align}
By combining these matrices, we can obtain the neutrino mass matrices $m_\nu$.
Among all combinations, the realistic patterns of $m_\nu$  are classified ~\cite{Barreiros:2018ndn}: 
\begin{align}
a:  \left[\begin{array}{ccc}\times& 0  &    \times \\ 0 & \times  & \times \\ \times &\times&\times   \\ \end{array}\right],\
b: \left[\begin{array}{ccc}  \times&  \times & 0 \\ \times & \times  & \times \\ 0 &\times&\times   \\ \end{array}\right],\
c: \left[\begin{array}{ccc}  \times&  \times & \times \\ \times & 0  & \times \\ \times &\times&\times   \\ \end{array}\right],\
d: \left[\begin{array}{ccc}  \times&  \times & \times \\ \times & \times  & \times \\ \times &\times&0   \\ \end{array}\right].\
\label{eq:neut-mat}
 \end{align}
These are one-zero textures.
Explicitly, these patterns are realized by the following combinations: 
 $a$ for $(T_{1,4},R_1)$, $b$ for $(T_{2,5},R_1)$, $c$ for $(T_{3,4},R_2)$ or $(T_{1,6},R_3)$ or $(U_{1,3},S)$ or $(U_{1},R_2)$
 or $(U_{3},R_3)$,
and $d$ for $(T_{5,6},R_2)$ or $(T_{2,3},R_3)$ or $(U_{2,4},S)$ or $(U_{2},R_2)$
 or $(U_{4},R_3)$.
 However since all the combinations including $U$ requires more Higgs doublets than those with $T_{1-6}$, we do not consider these cases.
The other combinations lead to the neutrino mass matrix $m_\nu$, which is not 
compatible with the experimental data.
Furthermore, all of the above patterns are compatible with the experiments 
for the inverted hierarchy, but not for the normal hierarchy.
Also obviously, one of neutrinos is massless.
For the above patterns of $m_\nu$, 
one finds the following relations~\cite{Barreiros:2018ndn}
\begin{align}
\frac{m_1}{m_2}&=-\frac{(U_{\rm PMNS}^*)_{i2}(U_{\rm PMNS}^*)_{j2} }{(U_{\rm PMNS}^*)_{i1} (U_{\rm PMNS}^*)_{j1}},\label{eq:cond1}\\
\frac{1}{1+r_\nu}&=\left|\frac{(U_{\rm PMNS}^*)_{i2}(U_{\rm PMNS}^*)_{j2} }{(U_{\rm PMNS}^*)_{i1} (U_{\rm PMNS}^*)_{j1}}\right|^2,
\quad r_\nu\equiv \frac{\Delta m^2_{21}}{\Delta m^2_{31}},\label{eq:cond2}
\end{align}
where {\it we can identify $\Delta m^2_{21}{+|\Delta m^2_{31}|}=m^2_{2}$ and $|\Delta m^2_{31}|=m^2_{1}$, since only inverted hierarchy is allowed for all the textures
by the current neutrino oscillation data.}
Moreover, $\cos\delta$ can be written in terms of observables and $r_\nu$ by solving Eq.(\ref{eq:cond1}) directly, 
while $\cos\alpha$ is also obtained  in terms of the same parameters of $\cos\delta$ by the fact that the imaginary part of Eq.(\ref{eq:cond2}) is zero.~\footnote{Neutrino mass eigenvalues are positive and real without loss of generality, because of reduced mass matrix.}



\begin{center} 
\begin{table}[tb]
\begin{tabular}{|c||c|c|c|c|c|c|c|c||c|c|c|c|c|c|c|c||}\hline\hline  
Fields & ~$L_{L_e}$~& ~$L_{L_\mu}$~ & ~$L_{L_\tau}$~& ~$e_{R}$~& ~$\mu_{R}$~ & ~$\tau_{R}$~ & ~$N_{R_1}$~& ~$N_{R_2}$~  & ~$H_{SM}$~& ~$H_1$~ &~$H_2$~ & ~$H_3$~ & ~$H_4$~ & ~$\varphi_1$~ & ~$\varphi_2$~ & ~$\varphi_3$~
\\\hline 
 $SU(2)_L$ & $\bm{2}$  & $\bm{2}$  & $\bm{2}$ & $\bm{1}$   & $\bm{1}$  & $\bm{1}$ & $\bm{1}$ & $\bm{1}$   & $\bm{2}$  & $\bm{2}$ & $\bm{2}$  & $\bm{2}$ & $\bm{2}$ & $\bm{1}$ & $\bm{1}$ & $\bm{1}$ \\\hline 
$U(1)_Y$ & $-\frac12$ & $-\frac12$ & $-\frac12$  & $-1$& $-1$ & $-1$  & $0$  & $0$   & $\frac12$ & $\frac12$  & $\frac12$ & $\frac12$ & $\frac12$ & $0$  & $0$ & $0$  \\\hline
 $U(1)_{\mu-\tau}$ & $0$ & $1$ & $-1$ & $0$ & $1$ & $-1$ & $n_1$ & $n_2$ & $0$ & $n_{1}$   & $n_{2}$-$1$ & $n_1$+$1$ & $n_2$+$1$ & -2$n_1$ & -2$n_2$& -$n_1$-$n_2$   \\\hline
\end{tabular}
\caption{Field contents of fermions and bosons
and their charge assignments under $SU(2)_L\times U(1)_Y\times U(1)_{\mu-\tau}$ in the neutrino to realize the one-zero neutrino textures $T_4$, where $n_1\neq n_2$, $n_1,n_2 \neq 0$ and 
$(n_{1},n_{2}\pm1,n_{1}+1)\neq \pm1,+2$.}
\label{tab:mu-tau}
\end{table}
\end{center}

\section{ Realizations of texture zeros} 
\label{sec:realization}

Here, we study realization by use of global $U(1)$ symmetry, $S_3$ and $D_4$ as well as 
$Z_N$.

\subsection{Abelian symmetries}
Here we consider a global $U(1)$ symmetry to realize predictive textures, where we fix the number of right-handed neutrinos to be two generations, i.e. $N_{R_{1,2}}$.
A flavor-dependent $U(1)$ symmetry in the lepton sector is useful to realize the diagonal mass matrix 
of the charged lepton sector.
That is, the  $U(1)_{\mu-\tau}$, $U(1)_{e-\mu}$ and $U(1)_{e-\tau}$ would be good candidates.
Here, let us study the realization of the Dirac mass texture $T_4$ by assuming 
the global $U(1)_{\mu-\tau}$ symmetry.\footnote{A gauged symmetry will be analyzed in elsewhere, since several phenomenologies are very different from the global one. {A comprehensive study has been done, {\it e.g.}, by ref.~\cite{Araki:2012ip} in which two-zero textures are realized, imposing two flavor dependent $U(1)$ gauge symmetries.}}
The assignment of $U(1)_{\mu - \tau}$ charges is shown in Table \ref{tab:mu-tau}.
We also assign  $U(1)_{\mu - \tau}$ charges, $n_1$ and $n_2$ to $N_{R_1}$ and $N_{R_2}$.
In order to realize Dirac neutrino mass terms, we have to introduce new $SU(2)_L$ doublet Higgs 
fields $H_i$, and their minimal number is four, i.e., $H_i$ ($i=1,2,3,4$).
Also, in order to realize the mass matrix $M_N$, we have to introduce singlet scalar fields, 
$\varphi_{1,2,3}$.
Here the charges $n_1,n_2$  should satisfy the condition,  $n_1\neq n_2$ and $n_1,n_2 \neq 0$  in order to realize 
the desired Dirac texture of $T_4$, and 
they should also satisfy 
$(n_{1},n_{2}\pm1,n_{1}+1)\neq \pm1$, $+2$ to forbid non-diagonal entries in the charged-lepton mass matrix.
Under these symmetries and fields, one can write renormalizable coupling terms in the Lagrangian as follows:
\begin{align}
-{\cal L}_{Lepton} &=
\sum_{\ell=e,\mu,\tau} y_{\ell} \bar L_{L_\ell} H_{SM} \ell_{R}
\nn\\
&
+ y_{D_1} \bar L_{L_e} \tilde H_1 N_{R_{1}} + y_{D_2} \bar L_{L_\mu} \tilde H_2 N_{R_{2}} + y_{D_3} \bar L_{L_\tau} \tilde H_3 N_{R_1}
+ y_{D_4} \bar L_{L_\tau} \tilde H_4 N_{R_2}\\
&+ y_{N_1} \bar N^C_{R_1} N_{R_1}\varphi_1
+ y_{N_2} \bar N^C_{R_2} N_{R_2}\varphi_2
+ y_{N_3} \bar N^C_{R_1} N_{R_2}\varphi_3
+ {\rm h.c.}, \label{eq:lag-lep}
\end{align}
where several dangerous Goldstone bosons (GBs)
can be evaded by introducing soft-breaking mass terms under $U(1)_{\mu-\tau}$ symmetry;
$m_{ij}^2 H^\dag_i H_j+{\rm h.c.}$ $i\neq j=$1-4.

After the spontaneous symmetry breaking,  the charged-lepton mass matrix and Dirac neutrino mass matrix are given by
\begin{align}
m_\ell&= \frac {v_H}{\sqrt{2}}
\left[\begin{array}{ccc}
y_e  & 0 & 0 \\ 
0 & y_\mu &  0 \\ 
0 & 0  & y_\tau  \\ 
\end{array}\right]
\equiv
\left[\begin{array}{ccc}
m_e & 0 & 0 \\ 
0 &m_\mu &  0 \\ 
0 & 0  & m_\tau \\ 
\end{array}\right]
,\\
m_D(T_4)&=\frac1{\sqrt{2}}
\left[\begin{array}{cc}
y_{D_1} v_{H_1} & 0  \\ 
0 & y_{D_2} v_{H_2}   \\ 
y_{D_2} v_{H_3} &  y_{D_4} v_{H_4}  \\
\end{array}\right]
\equiv
\left[\begin{array}{cc}
m_{D_1}  & 0 \\ 
0 & m_{D_2}  \\ 
m_{D_3} & m_{D_4}   \\ 
\end{array}\right], 
\end{align}
where $v_H$ and $v_{Hi}$ denote vacuum expectation values (VEVs) 
of the neutral components of $H_{SM}$  and $H_i$, respectively.
Then, the  $T_4$ pattern of the Dirac neutrino mass matrix in Ref.~\cite{Barreiros:2018ndn} is derived.
Also the right-handed neutrino mass matrix is given by
\begin{align}
M_N&=\frac1{\sqrt{2}}
\left[\begin{array}{cc}
y_{N_1} v_{\varphi_1} & y_{N_3} v_{\varphi_3}  \\ 
 y_{N_3} v_{\varphi_3}   &y_{N_2} v_{\varphi_2}   \\ 
\end{array}\right]
\equiv
\left[\begin{array}{cc}
M_1  & M_{12} \\ 
M_{12} & M_{2}  \\ 
\end{array}\right],
\end{align}
where $v_{\varphi_i}$ denote VEVs of $\varphi_i$. 
From the above equation, one straightforwardly finds each of texture $R_1$, $R_2$, and $R_3$ in absence of $\varphi_3$, $\varphi_1$, and $\varphi_2$.

We can realize the Dirac neutrino mass texture $T_1$ with the same charge assignment 
except replacing the charges of $H_1$ and $H_2$ such that 
$H_1$ and $H_2$ have $U(1)_{\mu -\tau}$ charges, $n_2$ and $n_1-1$.
Then, we can realize the Dirac neutrino mass,
\begin{align}
m_D(T_1)&=\frac1{\sqrt{2}}
\left[\begin{array}{cc}
0 & y_{D_1} v_{H_1}  \\ 
 y_{D_2} v_{H_2}  & 0 \\ 
y_{D_2} v_{H_3} &  y_{D_4} v_{H_4}  \\
\end{array}\right]
\equiv
\left[\begin{array}{cc}
0 & m_{D_1}  \\ 
 m_{D_2} & 0 \\ 
m_{D_3} & m_{D_4}   \\ 
\end{array}\right]. 
\end{align}

Similarly, the patterns, $T_5$ and $T_2$, are realized by $U(1)_{e-\mu}$ instead of $U(1)_{\mu-\tau}$.
Also  the patterns, $T_6$ and $T_3$, can be realized by use of $U(1)_{\tau-e}$ instead of $U(1)_{\mu-\tau}$.

Once any global $U(1)$ symmetries realize these predictive one-zero neutrino textures, discrete Abelian symmetries $Z_N$ are also possible in the same field contents, where $N\le 19$.

\subsection{Non-Abelian discrete symmetries}
Here we study the realization with non-Abelian discrete symmetries~\cite{Ishimori:2010au}.

\subsubsection{$S_3$ symmetry}
First of all, we study the $S_3$ symmetry, which is the minimal group in the non-Abelian discrete symmetries.
The irreducible representations of $S_3$ are the doublet 2, and the trivial singlet $1$ and 
the non-trivial singlet $1'$.
Here,  we use the real representation~\cite{Ishimori:2010au}\footnote{Note here that the complex representations cannot construct the diagonal mass matrix of charged lepton.}, and their products are expanded as 
\begin{align}
&\left[\begin{array}{c}
x_1 \\ 
 x_2 \\ 
\end{array}\right]_2
\otimes 
\left[\begin{array}{c}
y_1 \\ 
 y_2 \\ 
\end{array}\right]_2=
(x_1y_1+x_2y_2)_1\oplus (x_1y_2-x_2y_1)_{1'}\oplus 
\left[\begin{array}{c}
x_1y_2+x_2y_1 \\ 
 x_1y_1-x_2y_2 \\ 
\end{array}\right]_2,\\
&\left[\begin{array}{c}
x_1 \\ 
 x_2 \\ 
\end{array}\right]_2
\otimes 
(y')_{1'}=
\left[\begin{array}{c}
-x_2y' \\ 
 x_1y' \\ 
\end{array}\right]_2,\quad (x)_{1'}\otimes(y)_{1'}=(xy)_1.
\end{align}
We assign $(L_{L_\ell},\ell_R)$ $(\ell=e,\mu$) to the $S_3$ doublets $2$, and 
 $L_{L_\tau},\tau_R$ to the $S_3$ trivial singlets $1$.
In addition, we introduce four Higgs fields, which correspond to the $S_3$ doublet, $H_D\sim 2$, 
$S_3$ singlets, $H_1\sim1$, and $H_{2}\sim1'$. 
Then the renormalizable coupling terms of the charged-lepton sector are given by
\begin{align}
{\cal L}_\ell&=y_{\ell_1}[(\bar L_{L_e} H_{D_2}+\bar L_{L_\mu} H_{D_1})e_R +(\bar L_{L_e} H_{D_1} -\bar L_{L_\mu} H_{D_2})\mu_R]\nn\\
&+y_{\ell_2}(\bar L_{L_e} H_{D_1}+\bar L_{L_\mu} H_{D_2})\tau_R
+y_{\ell_3}(\bar L_{L_e} H_{1}e_R+\bar L_{L_\mu} H_{1}\mu_R)
+y_{\ell_4}\bar L_{L_\tau} (H_{D_1}e_R+ H_{D_2}\mu_R)\nn\\
&+y_{\ell_5}\bar L_{L_\tau} H_{1}\tau_R
+y_{\ell_6}(\bar L_{L_e} H_{2}\mu_R-\bar L_{L_\mu} H_{2} e_R)+{\rm h.c.}.
\end{align}
After the spontaneously electroweak symmetry breaking, the charged-lepton mass matrix can be found as
\begin{align}
m_\ell&= \frac {1}{\sqrt{2}}
\left[\begin{array}{ccc}
y_{\ell_1} v_{D_1}+y_{\ell_3} v_{1}  & y_{\ell_1} v_{D_1}+y_{\ell_6} v_{2}  & y_{\ell_2} v_{D_1} \\ 
y_{\ell_1} v_{D_2}-y_{\ell_6} v_{2}  & -y_{\ell_1} v_{D_2}+y_{\ell_3} v_{1}  &  y_{\ell_2} v_{D_2} \\ 
 y_{\ell_4} v_{D_1} &  y_{\ell_4} v_{D_2}  &  y_{\ell_5} v_{1}  \\ 
\end{array}\right],
\end{align}
where VEVs are denoted by $\langle H_i\rangle\equiv v_i/\sqrt2$ and $\langle H_{D_i}\rangle\equiv v_{D_i}/\sqrt2$ for $i=1,2$.
Once $\langle H_D\rangle=\langle H_2\rangle=0$, the diagonal charged-lepton mass matrix is realized;
\begin{align}
m_\ell&= \frac {1}{\sqrt{2}}
\left[\begin{array}{ccc}
y_{\ell_3} v_{1}  & 0  & 0\\ 
 0 & y_{\ell_3} v_{1}  &  0 \\ 
0 &0  &  y_{\ell_5} v_{1}  \\ 
\end{array}\right].
\end{align}
However, from the above mass matrix, one cannot reproduce the mass difference between the masses of electron and muon.
Thus, $S_3$ symmetry is not favorable.
\footnote{Note here that refs.~\cite{Gomez-Izquierdo:2018jrx, Gomez-Izquierdo:2017rxi} realize the appropriate charged-lepton mass matrix, by imposing an additional $Z_2$ symmetry.}


\subsubsection{$D_4$ symmetry}
\label{sec:D4}
Next, we investigate the $D_4$ flavor symmetry that is the next minimal group in the non-Abelian discrete symmetries.
The irreducible representations of $D_4$ symmetry are the doublet $2$, and the trivial singlet $1$, and 
three non-trivial singlets, $1',1'',1'''$.\footnote{The singlets, $1,1',1'',1'''$, correspond to 
$1_{++},1_{--},1_{+-},1_{-+}$ in Ref.~\cite{Ishimori:2010au}, respectively.}
Here, we also use the real representation, and their productions are shown in Appendix.
We assign $(L_{L_\ell},\ell_R)$ $(\ell=e,\mu$) to the $D_4$ doublets $2$, and 
 $L_{L_\tau},\tau_R$ to the $D_4$ trivial singlets $1$.
In addition, we introduce 6 Higgs fields, which correspond to all of the $D_4$ irreducible 
representations, $2$, $1,1',1',1'',1'''$,  that is, 
 $H_D\sim 2$, $H_1\sim1$,  $H_2\sim1'$,  $H_3\sim1''$,  $H_4\sim1'''$.
Then the renormalizable coupling terms of the charged-lepton sector are given by
\begin{align}
{\cal L}_\ell&=
y_{\ell_1}(\bar L_{L_e} H_{D_1}+\bar L_{L_\mu} H_{D_2})\tau_R
+y_{\ell_2}\bar L_{L_\tau} (H_{D_1}e_R+ H_{D_2}\mu_R)
+y_{\ell_3}(\bar L_{L_e} H_{1}e_R+\bar L_{L_\mu} H_{1}\mu_R)\nn\\
&
+y_{\ell_4}(\bar L_{L_e} H_{2}e_R-\bar L_{L_\mu} H_{2}\mu_R)
+y_{\ell_5}(\bar L_{L_e} H_{3}\mu_R+\bar L_{L_\mu} H_{3}e_R)
+y_{\ell_6}(\bar L_{L_e} H_{4}\mu_R-\bar L_{L_\mu} H_{4}e_R)\nn\\
&+y_{\ell_7}\bar L_{L_\tau} H_{1}\tau_R
+{\rm h.c.}.
\end{align}
After the spontaneously electroweak symmetry breaking, the charged-lepton mass matrix can be found as
\begin{align}
m_\ell&= \frac {1}{\sqrt{2}}
\left[\begin{array}{ccc}
y_{\ell_3} v_{1}+y_{\ell_4} v_{2}  & y_{\ell_5} v_{3}+y_{\ell_6} v_{4}  & y_{\ell_1} v_{D_1} \\ 
y_{\ell_5} v_{3}-y_{\ell_6} v_{4}  & y_{\ell_3} v_{1}-y_{\ell_4} v_{2}  &  y_{\ell_1} v_{D_2} \\ 
 y_{\ell_2} v_{D_1} &  y_{\ell_2} v_{D_2}  &  y_{\ell_7} v_{1}  \\ 
\end{array}\right],
\end{align}
where their VEVs are denoted by $\langle H_i\rangle\equiv v_i/\sqrt2$ ($i=1,\cdots,4$) and $\langle H_{D_i}\rangle\equiv v_{D_j}/\sqrt2$   for $j=1,2$.
Once $\langle H_D\rangle=\langle H_{2,3,4}\rangle=0$ and/or $y_{\ell_{1, 2, 5, 6}} = 0$, the diagonal charged-lepton mass matrix is realized;
\begin{align}
m_\ell&= \frac {1}{\sqrt{2}}
\left[\begin{array}{ccc}
y_{\ell_3} v_{1}+y_{\ell_4} v_{2}   & 0  & 0\\ 
 0 & y_{\ell_3} v_{1}-y_{\ell_4} v_{2}    &  0 \\ 
0 &0  &  y_{\ell_7} v_{1}   \\ 
\end{array}\right].
\end{align}
From the above equation, one can reproduce the mass difference between the masses of electron and muon.
Thus the $D_4$ flavor symmetry can be the minimal candidate to reproduce  the desired textures.
To realize the diagonal mass matrix of the charged lepton sector, 
we just need $H_1$ and $H_2$, but we do not need $H_D$ or $H_{3,4}$.

Next, let us explore the neutrino sector; Dirac and Majorana masses.
We classify the models by assigning systematically two right-handed neutrinos to 
two of the $D_4$ irreducible representations,  $2,1,1',1'',1'''$.

\noindent \underline{\it In the case of $(N_{R_1},N_{R_2}) \sim 2$},
the Majorana mass matrix is given by
\begin{align}
\label{eq:MN-2}
M_N=M \left[\begin{array}{cc}1  & 0\\ 0 & 1 \\ \end{array}\right],
\end{align}
where these two masses are degenerated. 
Then the Dirac neutrino  mass matrix is given by
\begin{align}
m_D&=\frac1{\sqrt{2}}
\left[\begin{array}{cc}
y_{D_1} v_{1} + y_{D_2} v_{2} & y_{D_3} v_{3} + y_{D_4} v_{4}   \\ 
y_{D_3} v_{3} - y_{D_4} v_{4}  & y_{D_1} v_{1} - y_{D_2} v_{2}  \\ 
y_{D_5} v_{D_1} &  y_{D_5} v_{D_2}  \\
\end{array}\right].
\end{align}
Hence one finds the desired Dirac mass matrix in the case of $\langle H_{3,4}\rangle=0$
\footnote{In case of $\langle H_{1,2}\rangle=0$, $T_1$ can be obtained. However, the electron and muon are massless.
Thus this case is ruled out.}
\begin{align}
\label{eq:MD-2}
m_D(T_4)=
\left[\begin{array}{cc}
  m_{D_1} &0  \\ 
0 & m_{D_2}     \\ 
m_{D_3}   &  m_{D_4}    \\
\end{array}\right].
\end{align}
For this realization, we need $H_D$, $H_1$ and $H_2$, but not $H_3$ or $H_4$.

Now, let us study the models, that $N_{R_1}$ and $N_{R_2}$ are assigned to 
two $D_4$ singlets.
If one assigns $N_{R_1}$ and $N_{R_2}$ into the same singlet representation under $D_4$, the Majorana mass matrix does not give any vanishing elements without imposing additional symmetries. 
Thus, we restrict ourselves to the models such that  $N_{R_1}$ and $N_{R_2}$ are assigned to 
$D_4$ singlets different from each other.

When we assign $N_{R_1}$ and $N_{R_2}$ into different $D_4$ singlets such as 
 $(N_{R_1},N_{R_2}) \sim (1,1')$, $ (1'',1''')$, etc. ,  
the Majorana mass matrix is give by
\begin{align}
M_N= \left[\begin{array}{cc}M_1  & 0\\ 0 & M_2 \\ \end{array}\right].
\end{align}
That is the $R_1$ form.

\noindent \underline{\it In the case of $(N_{R_1},N_{R_2}) \sim (1,1')$}, 
 the Dirac neutrino Yukawa mass matrix is given by
\begin{align}
m_D &=\frac1{\sqrt{2}}
\left[\begin{array}{cc}
y_{D_1} v_{D_1} & y_{D_2} v_{D_1}    \\ 
 y_{D_1} v_{D_2}   & - y_{D_2} v_{D_2}  \\ 
y_{D_3} v_{1} &  y_{D_4} v_{2}  \\
\end{array}\right].\
\end{align}
This  form cannot clearly  reproduce any types of desired Dirac mass matrices,
since $y_{D_1}$ and $y_{D_2}$ are located in the same column of upper $(2 \times 2)$ matrix.
When we assign $(N_{R_1},N_{R_2}) \sim (1'',1''')$, we obtain a similar result.
Then, these two cases are not favorable, but the other cases are favorable.


\noindent \underline{\it In the case of $(N_{R_1},N_{R_2}) \sim (1(1'),1''(1'''))$},
 the Dirac neutrino mass matrix is given by
\begin{align}
m_D&=\frac1{\sqrt{2}}
\left[\begin{array}{cc}
y_{D_1} v_{D_1} & y_{D_2} v_{D_2}    \\ 
\pm y_{D_1} v_{D_2}   & \pm y_{D_1} v_{D_1}  \\ 
y_{D_3} v_{1} &  y_{D_4} v_{3(4)}  \\
\end{array}\right],
\end{align}
where $"+"$ and $"-"$ in the (2,1) component corresponds to $N_{R_1}\sim 1$ and $N_{R_1}\sim 1'$, respectively, and 
$"+"$ or $"-"$ in the (2,2) component and $v_3$ and $v_4$ in the (3,3) component correspond to $N_{R_2}\sim 1''$ and  
$N_{R_2}\sim 1'''$, respectively.
One straightforwardly finds the desired Dirac mass matrices $T_1$ and $T_4$ in the cases with $\langle H_{D_1}\rangle=0$ and $\langle H_{D_2}\rangle=0$,
respectively. 
For example, in the case of  $(N_{R_1},N_{R_2}) \sim (1,1'')$, we need $H_D$, $H_{1,2,3}$, but not $H_4$.

In order to obtain $T_{2,3,5,6}$, one straightforwardly finds them by reassigning the fields of the SM leptons.
For example, once we assign $(L_{L_e},L_{L_\tau})\sim(e_R,\tau_R)\sim 2$, and $(L_{L_\mu},\mu_R)\sim1$,
then one finds $T_2$ or $T_5$.
On the other hand, when we assign $(L_{L_\mu},L_{L_\tau})\sim(\mu_R,\tau_R)\sim 2$, and $(L_{L_e},e_R)\sim1$,
then one finds $T_3$ or $T_6$.

To summarize results in this section, one can realize the desired textures by $D_4$, but not by $S_3$.
Indeed, the $D_4$ flavor symmetry is interesting from the viewpoints of both high energy physic 
\cite{Kobayashi:2006wq,Kobayashi:2004ya,Ko:2007dz,Abe:2009vi,Abe:2009uz,
BerasaluceGonzalez:2012vb,Marchesano:2013ega,Abe:2014nla,Kobayashi:2018rad}.
and bottom-up model building approach \cite{Grimus:2003kq,Grimus:2004rj,Ishimori:2008gp,Adulpravitchai:2008yp,
Hagedorn:2010mq,Meloni:2011cc}.
Similarly, we can discuss realization by using other non-Abelian discrete flavor symmetries.
Also we can realize the desired textures by Abelian symmetries, $U(1)$ and $Z_N$.
We need more Higgs fields in the Abelian models than 
the $D_4$ models.
Thus, the $D_4$ flavor symmetry is useful to realize the desired textures.
{Note here that the textures $c$ and $d$ in Eq.~(\ref{eq:neut-mat}) cannot be realized by $D_4$ symmetry, because $M_N$ is diagonal ($R_1$ form).~\footnote{If an additional symmetry is introduced in basis of $(N_{R_{1}},N_{R_{2}})\sim1$ under $D_4$ symmetry, $c$ and $d$ can be realized.
But this is beyond our scope.}  }
In the next section, we propose a concrete model with the $D_4$ flavor symmetry.

 \begin{widetext}
\begin{center} 
\begin{table}[t]
\begin{tabular}{|c||c|c|c|c|c|c||c|c|c|c|c|c|c|c||c|c|}\hline\hline  
Fields & ~$L_{L_\ell}$~ & ~$L_{L_\tau}$~& ~$\ell_{R}$~ & ~$\tau_{R}$~ & ~$N_{R_i}$~ & ~$N_{R_\tau}$~ & ~$H$~& ~$H_2$~& ~$\eta_1$~ &~$\eta_{1'}$~ & ~$\eta_{D}$~ & ~$\varphi_{8}$~ & ~$\varphi'_{8}$~ & ~$\varphi_{10}$~ & ~$\zeta$~ & ~$\varphi_{2}$
\\\hline 
 $SU(2)_L$ & $\bm{2}$  & $\bm{2}$  & $\bm{1}$ & $\bm{1}$ & $\bm{1}$  & $\bm{1}$  & $\bm{2}$ & $\bm{2}$ & $\bm{2}$ & $\bm{2}$  & $\bm{2}$ & $\bm{1}$& $\bm{1}$& $\bm{1}$ & $\bm{2}$ & $\bm{1}$   \\\hline 
$U(1)_Y$ & $-\frac12$ & $-\frac12$  & $-1$ & $-1$  & $0$ & $0$  & $\frac12$  & $\frac12$ & $\frac12$  & $\frac12$ & $\frac12$ & $0$ & $0$ & $0$ & $\frac12$ & $0$    \\\hline
 $U(1)_{B-L}$ & $-1$ & $-1$  & $-1$ & $-1$ & $-4$ & $5$  & $0$ & $0$ & $-3$   & $-3$   & $-3$  & $8$   & $8$  & $10$  & $-6$ & $2$  \\\hline
 $D_{4}$ & $2$ & $1$ & $2$ & $1$ & $2$ & $1$  & $1$ & $1'$ & $1$   & $1'$   & $2$  & $1$   & $1'$  & $2$  & $1$ & $1$  \\\hline
\end{tabular}
\caption{Field contents of fermions and bosons
and their charge assignments under $SU(2)_L\times U(1)_Y\times U(1)_{B-L}\times D_4$ in the neutrino and Higgs sector, where $\ell=e,\mu$ is flavor index.}
\label{tab:1}
\end{table}
\end{center}
\end{widetext}

\section{A concrete model in $D_4$ symmetry}
\label{sec:model}

Here, we study a concrete model based on the $D_4$ symmetry.
First, we explain our setup.
Basically, our model corresponds to the scenario, where $N_{R_{1,2}}$ are assigned to 
the $D_4$ doublet in Sec.~\ref{sec:D4}.
In addition, we also introduce the third right-handed neutrino $N_{R_3}$, 
but arrange it such that $N_{R_3}$ has no Dirac mass term with left-handed neutrino and 
no Majorana mass terms with $N_{R_{1,2}}$.
For such a purpose, 
we assume additional $U(1)$ gauge symmetry, that is, $U(1)_{B-L}$.
Its charge assignment is the same as the conventional one except the right-handed neutrino sector.
For the right-handed neutrino sector, we assign 
 $U(1)_{B-L}$ charges, $-4, -4,5$ to $N_{R_{1,2,3}}$,  respectively.
That is the so-called alternative $U(1)_{B-L}$ \cite{Montero:2007cd,Ma:2014qra, Singirala:2017see, Nomura:2017vzp,Nomura:2017jxb,Nomura:2017kih,Geng:2017foe}.
All gauge anomalies are canceled with this choice.
In the boson sector,
we introduce several new bosons $H_2,\eta_{1,1',D}, \varphi_{2,8}, \varphi'_{8},\zeta$ in addition to the SM Higgs $H$, where $H$ gives the masses for the quark sector and the charged lepton sector, while $H_2$ gives mass difference between electron(positron) and muon(antimuon).
Here their VEVs are symbolized by $\langle H\rangle\equiv v_H,\langle H_2\rangle\equiv v'_H,\langle\eta_{1,1',D}\rangle\equiv v_{\eta,\eta'\eta_D}, \langle\varphi_{2,8}\rangle\equiv v_{\varphi_2,\varphi_8}, \langle\varphi'_{8}\rangle\equiv v_{\varphi'_8}, \langle\zeta\rangle\equiv v_{\zeta}$.
Also $\eta$ and $\varphi_8$ respectively provide the Dirac and right-handed neutrino masses, 
$\eta'$ and $\varphi'_{8}$ respectively provide the difference between the (1-1) and (2-2) elements of $m_D$ and $M_N$,
and $\eta_D$ gives the masses for the third row of Dirac mass matrix.
$\zeta$ and $\varphi_2$ play a role in evading  dangerous GBs due to accidental symmetries in the scalar potential.  
The $D_4$ symmetry assures diagonal mass matrices for charged leptons and right-handed neutrinos,
and $U(1)_{B-L}$ plays a role in restricting $(2\times 2)$ mass matrix for right-handed neutrinos which contribute to active neutrino masses.
In addition, our $U(1)_{B-L}$ charge assignment makes $N_{R_3}$ stable and it can be a DM candidate. 
All the field contents and their charge assignments are shown in Table.~\ref{tab:1}.
Under these contents with symmetries, one can write  renormalizable Yukawa coupling terms and the Higgs potential as follows:~\footnote{We show valid multiplication rules for $D_4$ in Appendix.}
\begin{align}
-{\cal L}_{Lepton} =&
y_{\ell} (\bar L_{L_e} e_{R}+\bar L_{L_\mu} \mu_{R})H + y'_{\ell}  (\bar L_{L_e} e_{R} - \bar L_{L_\mu} \mu_{R}) H_2 
+y_{\tau} \bar L_{L_\tau} \tau_{R} H\nn\\
&
+ y_{D} (\bar L_{L_e} N_{R_e} + \bar L_{L_\mu} N_{R_\mu} ) \tilde\eta_1
 + y'_{D}  (\bar L_{L_e} N_{R_e}  - \bar L_{L_\mu} N_{R_\mu} )  \tilde\eta'_1\nn\\
& +y_{D_3} \bar L_{L_\tau} (N_{R_e} \tilde\eta_{D_1} +N_{R_\mu} \tilde\eta_{D_2} )\\
&+ y_{N} (\bar N^C_{R_e} N_{R_e} + \bar N^C_{R_\mu} N_{R_\mu})\varphi_8 
+ y'_{N} (\bar N^C_{R_e} N_{R_e} - \bar N^C_{R_\mu} N_{R_\mu})\varphi'_8 
+ {\rm h.c.},\nn\\
V =&
\lambda_{1,1',D} (\zeta^\dag \eta_{1,1',D})(H^\dag \eta_{1,1',D}) + \lambda'_1 (\zeta^\dag \eta_1)(H_2^\dag \eta_{1'})
+ \lambda'_D (\zeta^\dag \eta_D)(H_2^\dag \eta_{D})\nn\\
& 
+ \lambda_0 (\zeta^\dag H) \varphi^*_8\varphi_2
  + {\rm h.c.},
\label{eq:lag-lep}
\end{align}
where 
$V$ is the Higgs potential with non-trivial terms. These nontrivial terms forbid dangerous GBs arising from isospin doublets that spoil the model. 
In our model, we have two GBs that can be identified with CP-odd bosons of $\varphi_2$ and $\varphi_8(\varphi'_8)$.
\footnote{In addition, one has to introduce soft breaking terms of $D_4$ symmetry in order to forbid accidental symmetries that also induce dangerous GBs. The breaking patterns are given by ref.~\cite{Ishimori:2010au}, and any patterns are fine because it does not affect our model. Thus we do not discuss this issue further. }

\subsection{Lepton sector}
The resulting mass matrices are give by
\begin{align}
m_\ell&= \frac1{\sqrt{2}}
\left[\begin{array}{ccc}
y_\ell v_H +y'_\ell v'_H & 0 & 0 \\ 
0 & y_\ell v_H - y'_\ell v'_H &  0 \\ 
0 & 0  & y_\tau v_H \\ 
\end{array}\right]
\equiv
\left[\begin{array}{ccc}
m_e & 0 & 0 \\ 
0 &m_\mu &  0 \\ 
0 & 0  & m_\tau \\ 
\end{array}\right]
,\\
m_D&=\frac1{\sqrt{2}}
\left[\begin{array}{cc}
y_D v_{\eta} +y'_D v_{\eta'} & 0  \\ 
0 &y_D v_\eta - y'_D v_{\eta'}   \\ 
y_{D_3} v_{\eta_{D_1}} &  y_{D_3} v_{\eta'_{D_2}}  \\
\end{array}\right]
\equiv
\left[\begin{array}{cc}
m_{D_1}  & 0 \\ 
0 & m_{D_2}  \\ 
m_{D_3} & m_{D_4}   \\ 
\end{array}\right], \\
M_N&=\frac1{\sqrt{2}}
\left[\begin{array}{cc}
y_N v_{\varphi_8} +y'_N v_{\varphi'_8} & 0  \\ 
0 & y_N v_{\varphi_8} - y'_N v_{\varphi'_8}   \\ 
\end{array}\right]
\equiv
\left[\begin{array}{cc}
M_1  & 0 \\ 
0 & M_{2}  \\ 
\end{array}\right].
\end{align}
The above neutrino Dirac mass matrix $m_D$ corresponds to Eq.~(\ref{eq:MD-2}).
Also the above Majorana mass matrix $M_N$ basically corresponds to Eq.~(\ref{eq:MN-2}).
However, since there are two fields $\varphi_8$ and $\varphi'_8$, then we obtain 
$M_1 \neq M_2$.
Then, we can obtain 
\begin{align}
m_\nu&\approx
\left[\begin{array}{ccc}
\frac{m_{D_1}^2}{M_1} & 0 & \frac{m_{D_1}m_{D_3}}{M_1}  \\ 
0 & \frac{m_{D_2}^2}{M_2} & \frac{m_{D_2}m_{D_4}}{M_2}  \\ 
\frac{m_{D_1}m_{D_3}}{M_1}  & \frac{m_{D_2}m_{D_4}}{M_2}  &  \frac{m_{D_3}^2}{M_1} + \frac{m_{D_4}^2}{M_2}  \\ 
\end{array}\right], \label{eq:patt-a_mat}
\end{align}
which corresponds to the pattern $a$ in Eq.~(\ref{eq:neut-mat}).
%
%
%
%
%
%
%
Applying the discussion in Sec.~\ref{sec:review} to our model, we find 
\begin{align}
\frac{m_1}{m_2}=-\frac{(U_{\rm PMNS}^*)_{12}(U_{\rm PMNS}^*)_{22} }{(U_{\rm PMNS}^*)_{11} (U_{\rm PMNS}^*)_{21}}.
\label{eq:relation}
\end{align}
Therefore, one obtains two relations from the above relation:
{\begin{align}
\cos\delta&=\frac{[s^4_{12}(1+r_\nu)-c^4_{12}]s^2_{23}s^2_{13}+r_\nu c^2_{23}s^2_{12}c^2_{12}}
{2[s^2_{12}(1+r_\nu)+c^2_{12}]s_{12}c_{12}s_{23}c_{23}s_{13}},\label{eq:delta}\\
\cos\alpha&=\frac{-[s^4_{12}(1+r_\nu)+c^4_{12}]s^2_{23}s^2_{13}+(2+r_\nu)c^2_{23}s^2_{12}c^2_{12}}
{2\sqrt{(1+r_\nu}(c^2_{23}+s^2_{23}s^2_{13})s^2_{12}c^2_{12}}, 
\label{eq:alpha_neg}
\end{align}
where Eq.(\ref{eq:delta}) is derived by solving Eq.(\ref{eq:relation}) directly, while Eq.(\ref{eq:alpha_neg}) is obtained by the fact that the imaginary part of Eq.(\ref{eq:relation}) is vanishing.
Applying the current neutrino oscillation data~\cite{Forero:2014bxa},
we find some predictions.
In fig.~\ref{fig:a-d}, we show the allowed region between $\alpha/\pi$ and $\delta/\pi$ and
and it suggests as follows; $0.075\lesssim \alpha/\pi\lesssim0.15$ and $0.49\lesssim\delta/\pi\lesssim0.52$ at 3 $\sigma$ confidential level (CL) (blue region),
 $0.105\lesssim \alpha/\pi\lesssim0.13$ and $0.50\lesssim\delta/\pi\lesssim0.51$ at 1 $\sigma$ CL (red region), and $(\alpha/\pi,\delta/\pi)\approx(0.11,0.51)$ at best fit value (BF) (black dot).

}

{
\begin{figure}[t]
\centering
\includegraphics[width=10cm]{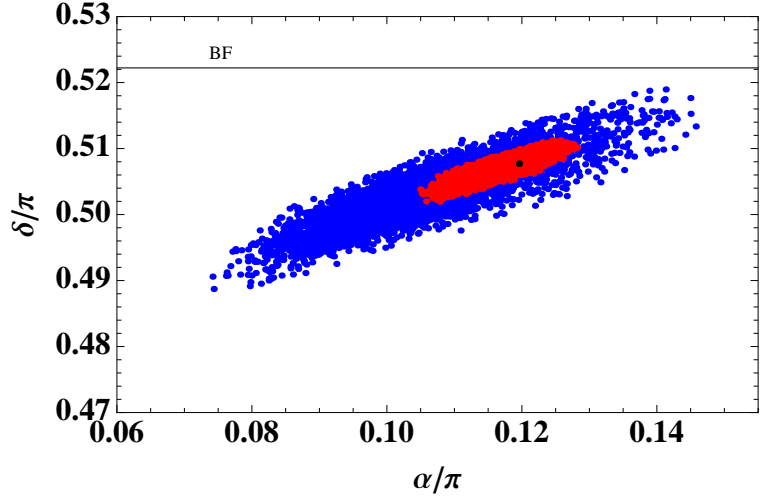}
\caption{{
Allowed region between $\alpha/\pi$ and $\delta/\pi$ to satisfy the current neutrino oscillation data. 
Also the blue, red, and black regions respectively represent predictions in light of the experimental input results at 3 $\sigma$ CL, 1 $\sigma$ CL, and BF. Here the black horizontal line presents the best fit value (BF).}}
\label{fig:a-d}
\end{figure}

{\it Consistency check}:
 Replacing $a\equiv\frac{m_{D_1}^2}{M_1} $,
 $b\equiv \frac{m_{D_2}^2}{M_2}$, $r_{31}\equiv m_{D_3}/m_{D_1}$, $r_{42}\equiv m_{D_4}/m_{D_2}$
Eq.~(\ref{eq:patt-a_mat}) can be rewritten in terms of four parameters as follows:
\begin{align}
m_\nu&\approx
\left[\begin{array}{ccc}
a & 0 &a r_{31}  \\ 
0 &b &b r_{42}  \\ 
a r_{31}  & b r_{42} & a r_{31}^2 +  b r_{42}^2   \\ 
\end{array}\right]. \label{eq:patt-a_rewrit}
\end{align}
It implies that $(m_{\nu})_{33}$ component is uniquely fixed once $a,b,r_{31},r_{42}$ are determined by experimental values.
While experimental value of $(m_{\nu})_{33}$; $((m_{\nu}^{exp})_{33}\equiv)[U^*_{\rm PMNS}{\rm diag}(m_1,m_2,m_3)U^\dag_{\rm PMNS}]_{33}$, is independently determined by experimental result, too.
In fig.~\ref{fig:33}, we show the allowed region between $|(m_{\nu})_{33}^{exp}|$ and $|(m_{\nu})_{33}|$, where the red line represents $|(m_{\nu})_{33}^{exp}|=|(m_{\nu})_{33}|$.
It suggests the theoretical consequence is in favor of the experimental result that is consistent with the original paper \cite{Barreiros:2018ndn}.


\begin{figure}[t]
\centering
\includegraphics[width=10cm]{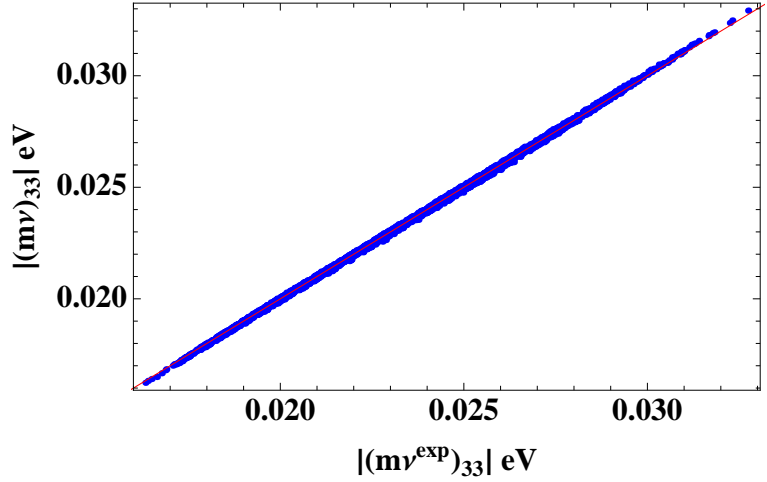}
\caption{{   
Allowed region between $|(m_{\nu})_{33}^{exp}|$ and $|(m_{\nu})_{33}|$, where the red line represents $|(m_{\nu})_{33}^{exp}|=|(m_{\nu})_{33}|$.}}
\label{fig:33}
\end{figure}

}

\subsection{Phenomenology}

In this subsection, we discuss phenomenology of the model such as collider physics and dark matter physics.
At the LHC $Z'$ can be produced as it couples to the SM quarks, and can decay into the SM leptons providing clear di-lepton signal. 
On the other hand the signatures from exotic scalar bosons are more complicated containing more particles in final states and their branching ratios depend on the parameters in the scalar potential 
so that we have less predictability, although they can be also produced via $Z'$ interaction and through electroweak interaction if an exotic scalar boson comes from iso-doublet. 
Thus we focus on $Z'$ production in $s$-channel followed by decay mode of $Z' \to \ell^+ \ell^-$ and estimate the constraints for new gauge coupling constant and mass of $Z'$.
Then dark matter relic density is briefly discussed taking into account the constraint for $Z'$ interaction.
 
\subsubsection{Collider physics and constraints}
 \begin{widetext}
\begin{center} 
\begin{table}[t]
\begin{tabular}{|c|cccccccccccc|}\hline\hline  
 & ~$\ell^+_i \ell^-_i$~ & ~$\bar \nu_i \nu_i$~ & ~$\bar q_i q_i$~ & ~$N_{R_{1,2}}$~ & ~$N_{R_3}$~ & ~$\eta_1^* \eta_1$~ & ~$\eta_{1'}^* \eta_{1'}$~ & ~$\eta_D^* \eta_D$~ & ~$\varphi_8^* \varphi_8$~ 
 & ~$\varphi'^*_8 \varphi'_8$~ & ~$\varphi^*_{10} \varphi_{10}$~ & ~$\zeta^* \zeta$~ \\ \hline 
 case (1) & $0.15$ & $0.077$ & $0.051$ & $0.0$ & $0.0$ & $0.0$ & $0.0$ & $0.0$ & $0.0$ & $0.0$ & $0.0$ & $0.0$  \\
 case (2) & $0.073$ & $0.037$ & $0.024$ & $0.15$ & $0.23$ & $0.0$ & $0.0$ & $0.0$ & $0.0$ & $0.0$ & $0.0$ & $0.0$ \\
 case (3) & $0.0076$ & $0.0038$ & $0.0025$ & $0.015$ & $0.024$ & $0.034$ & $0.034$ & $0.068$ & $0.12$ & $0.12$ & $0.19$ & $0.068$ \\ \hline
\end{tabular}
\caption{Branching ratios for $Z'$ decay in cases: (1) $m_{Z'} < 2 m_{N_{R_i}}$ and $m_{Z'} < 2 m_{\Phi}$; (2) $m_{Z'} > 2 m_{N_{R_i}}$ and $m_{Z'} < 2 m_{\Phi}$; 
(3) $m_{Z'} > 2 m_{N_{R_i}}$ and $m_{Z'} > 2 m_{\Phi}$ where we ignored dependence on final state mass assuming $m_{N_{R_i}, \Phi}^2 \ll m_{Z'}^2$ if kinematically allowed in case (2) and (3). For exotic scalar modes, BRs for all components are summed up. }
\label{tab:BRs}
\end{table}
\end{center}
\end{widetext}
Here we explore collider physics focusing on $Z'$ boson and provide constraints for its mass and gauge coupling constant.
The relevant gauge interactions are given by
\begin{align}
\mathcal{L}_{\rm int} =& g_{BL} Z'_\mu \biggl[ \frac{1}{3} \bar Q_L \gamma^\mu Q_L + \frac{1}{3} \bar u_R \gamma^\mu u_R + \frac{1}{3} \bar d_R \gamma^\mu d_R - \bar L \gamma^\mu L - \bar e_R \gamma^\mu e_R   \nonumber \\
& \qquad \qquad + \frac{1}{2} Q^{B-L}_{N_{R_i}} \bar N_{R_i} \gamma^\mu \gamma^5 N_{R_i} + Q^{B-L}_{\Phi} (\partial^\mu \Phi^* \Phi - \Phi^* \partial^\mu \Phi ) \biggr], 
\end{align}
where flavor indices for the SM fermions are omitted and $\Phi = \{\eta_1, \eta_{1'}, \eta_D, \varphi_8, \varphi'_8, \varphi_{10}, \zeta \}$; note that $\varphi_2$ is not included here since we assume its CP-odd component is Nambu-Goldstone boson absorbed by $Z'$.
The mass of $Z'$ is given by $m_{Z'} = g_{BL} \sqrt{\sum_{\Phi_{BL}} (Q^{B-L}_{\Phi} v_{\Phi_{BL}})^2 }$ where $\Phi_{BL}$ and $v_{\Phi_{BL}}$ indicate scalar field with $B-L$ charge $Q^{B-L}_{\Phi}$ and its VEV respectively.  
The partial decay widths of $Z'$ are estimated as 
\begin{align}
\Gamma_{Z' \to \bar f_{SM} f_{SM}} &= \frac{(Q^{B-L} g_{BL})^2 }{12 \pi} m_{Z'} \left( 1 - \frac{4 m_{f_{SM}}^2}{m_{Z'}^2} \right)^{\frac32}, \nonumber \\
\Gamma_{Z' \to \bar N_{R_i} N_{R_i}} &= \frac{(Q^{B-L}_{N_{R_i}} g_{BL})^2}{96 \pi} \left( 1 - \frac{4 m_{N_{R_i}}^2 }{m_{Z'}^2} \right)^{\frac32}, \nonumber \\
\Gamma_{Z' \to \Phi_1 \Phi_2} &= \frac{(Q_{\Phi}^{B-L} g_{BL})^2 }{48 \pi} m_{Z'} \lambda^{\frac12}(m_{Z'}, m_{\Phi_1}, m_{\Phi_2}) \left[ 1 - \frac{2 (m_{\Phi_1}^2 + m_{\Phi_2}^2)}{m_{Z'}^2} + \frac{(m_{\Phi_1}^2 - m_{\phi_2}^2)}{m_{Z'}^4} \right], \nonumber \\
\lambda (m_{Z'}, m_{\Phi_1}, m_{\Phi_2} ) &= 1 + \frac{m_{\Phi_1}^4}{m_{Z'}^4} + \frac{m_{\Phi_2}^4}{m_{Z'}^4} - 2 \frac{m_{\Phi_1}^2 m_{\Phi_2}^2}{m_{Z'}^4} - 2 \frac{m_{\Phi_1}^2}{m_{Z'}^2} - 2 \frac{m_{\Phi_2}^2}{m_{Z'}^2}, \label{eq:widths}
\end{align}
where $f_{SM}$ denotes the SM fermions and $\{\Phi_1, \Phi_2 \}$ indicate components of $\Phi$.
We estimate branching ratios (BRs) for $Z'$ decay in cases: (1) $m_{Z'} < 2 m_{N_{R_i}}$ and $m_{Z'} < 2 m_{\Phi}$; (2) $m_{Z'} > 2 m_{N_{R_i}}$ and $m_{Z'} < 2 m_{\Phi}$; 
(3) $m_{Z'} > 2 m_{N_{R_i}}$ and $m_{Z'} > 2 m_{\Phi}$, where $m_{\Phi}$ represents exotic scalar mass assuming they are mostly the same scale.
In TABLE.~\ref{tab:BRs}, we show the BRs for $Z'$ decay where we ignored dependence on final state mass assuming $m_{N_{R_i}, \Phi}^2 \ll m_{Z'}^2$ if kinematically allowed in cases (2) and (3) for simplicity.
We find that $BRs$ for the SM fermions are significantly suppressed when all exotic scalar modes are open.

\begin{figure}[t]
\centering
\includegraphics[width=8cm]{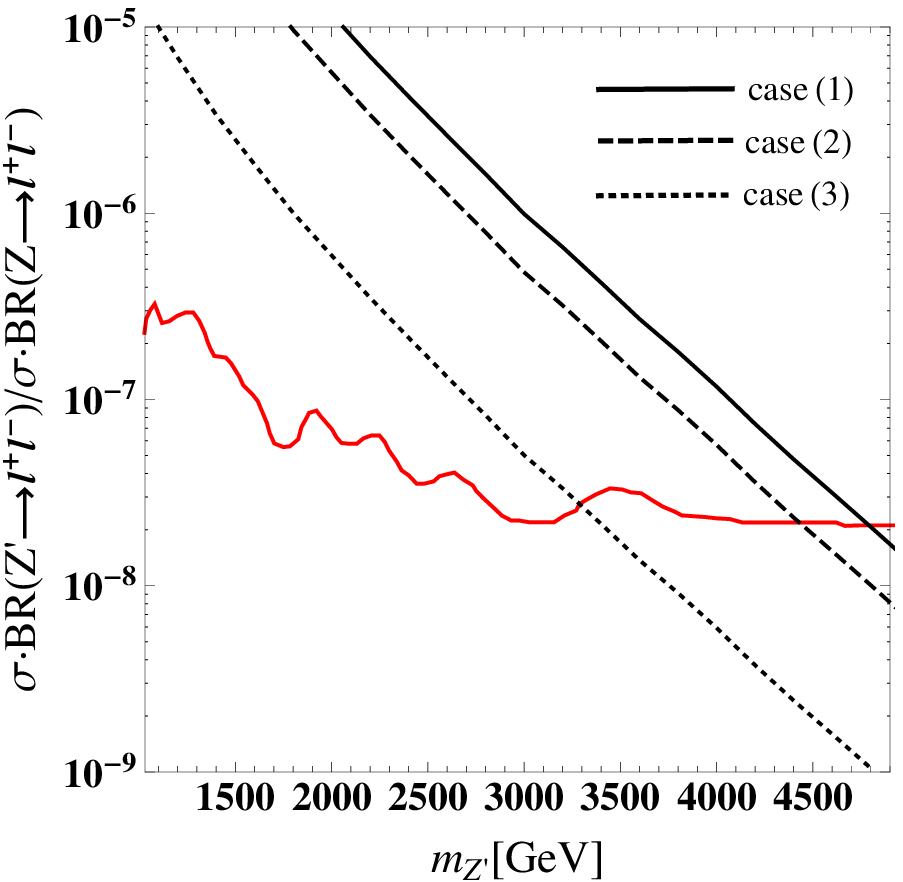}
\includegraphics[width=8cm]{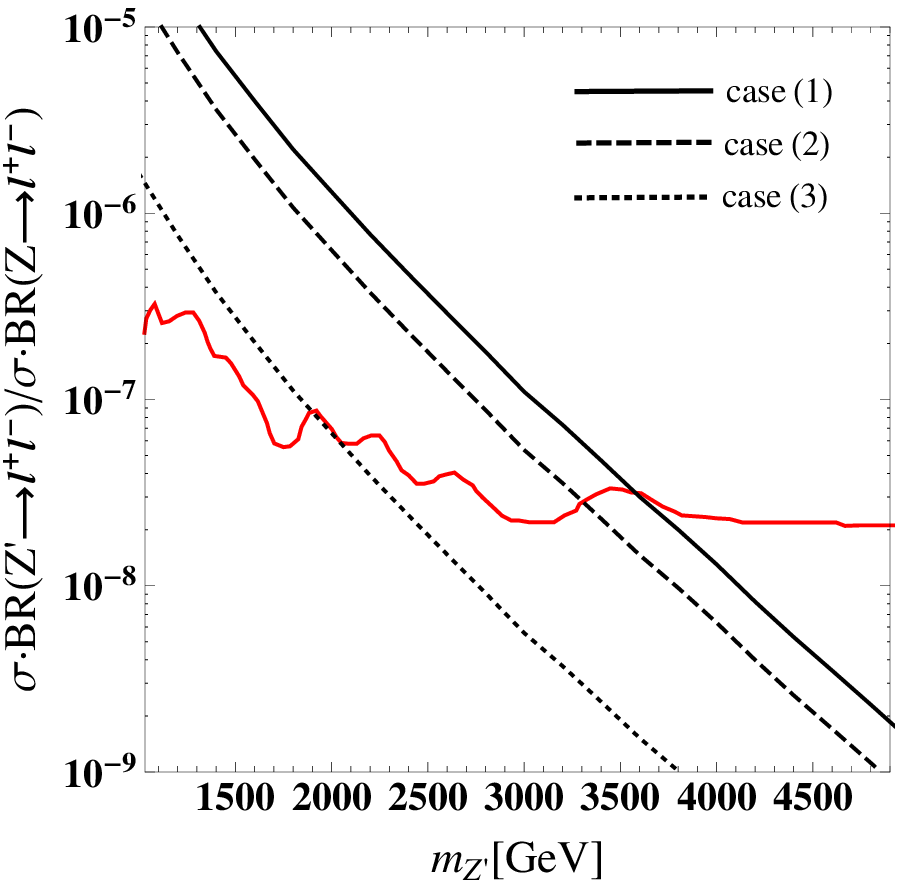}
\caption{The ratio between $\sigma \cdot BR (pp \to Z' \to \ell^+ \ell^-)$ and $\sigma \cdot BR (pp \to Z \to \ell^+ \ell^-)$ where $\ell = e, \mu$ and the red curve indicates the experimental constraints {which corresponds to 95$\%$ confidence level observed limit}. The left(right) plot corresponds to $g_{BL} = 0.3(0.1)$.}
\label{fig:Zp}\end{figure}

Then we discuss constraint on $g_{BL}$ from the LHC experiments for three cases above.
Our $Z'$ boson is produced via $Z' \bar q q$ coupling and the production cross section is estimated using {\tt CalcHEP 3.6}~\cite{Belyaev:2012qa} implementing relevant interactions.
The most stringent constraint comes from the process $pp \to Z' \to \ell^+ \ell^- (\ell =e, \mu)$ and we estimate the corresponding cross section for each case.
In Fig.~\ref{fig:Zp}, we compare ratio between $\sigma \cdot BR (pp \to Z' \to \ell^+ \ell^-)$ and $\sigma \cdot BR (pp \to Z \to \ell^+ \ell^-)$ in our model with the experimental constraints {corresponding to 95$\%$ confidence level (CL) observed limit}
 indicated by red curve~\cite{CMS:2018wsn} where solid, dashed and dotted curve correspond to cases (1), (2) and (3) respectively, and we apply $g_{BL} = 0.3(0.1)$ in left(right) plots.
Thus lower limit of mass $Z'$ is relaxed when the exotic scalar modes of $Z'$ decay are kinematically allowed: the lower limit of $m_{Z'}$ is around $3300(2000)$ GeV for $g_{BL} = 0.3(0.1)$ {in 95$\%$ CL}.
For case (3), the $Z'$ boson dominantly decays into exotic scalar bosons which further decay into SM particles via gauge interaction and/or couplings in the scalar potential providing multi-particle final states.
The detailed analysis of the scalar modes is beyond the scope of our analysis.

\subsubsection{Dark matter} 
In this subsection we discuss a dark matter candidate; $X_{R}\equiv N_{R_3}$, whose stability is assured by the $U(1)_{B-L}$ symmetry with alternative charge assignment for the SM singlet fermions.
Here, let us assume any contributions from the Higgs mediating interaction are negligibly small 
so as to avoid the constraints from direct detection searches as LUX~\cite{Akerib:2016vxi}, XENON1T~\cite{Aprile:2017iyp}, and PandaX-II~\cite{Cui:2017nnn}.
Then DM annihilation processes are dominated by the gauge interaction with $Z'$ and GB $\alpha_G\equiv z_{\varphi_{10}}$ mainly originated from $\varphi_{10}$,
and  their relevant Lagrangian in basis of mass eigenstate is found to be {
\begin{align}
-{\cal L} \supset &
\frac12 Q^X_{BL} g_{BL} \bar X\gamma^\mu \gamma_5 X Z'_\mu 
+i \frac{ M_{X} }{v_{\varphi_{10}}}  \bar X P_R X  \alpha_{G}+{\rm c.c.},  \label{eq:dmint}
\end{align}
where $Q^X_{BL}=5$, $M_X\equiv  y_{N_3}  v_{\varphi_{10}}/\sqrt{2}$, $v_{\varphi_{10}}<<v_{\varphi_{2}}$.
Here we require $Z'$ mass and gauge coupling $g_{BL}$ to satisfy the relation $g_{BL}/m_{Z'}\lesssim 1/(6.9\ {\rm TeV})$ from LEP experiment~\cite{Schael:2013ita} as well as the constraints from the LHC experiments as discussed in the previous subsection.
The relic density of DM is then given by~\cite{Griest:1990kh, Edsjo:1997bg}
\begin{align}
&\Omega h^2
\approx 
\frac{1.07\times10^9}{\sqrt{g_*(x_f)}M_{Pl} J(x_f)[{\rm GeV}]},
\label{eq:relic-deff}
\end{align}
where $g^*(x_f\approx25)$ is the degrees of freedom for relativistic particles at temperature $T_f = M_X/x_f$, $M_{Pl}\approx 1.22\times 10^{19}$ GeV,
and $J(x_f) (\equiv \int_{x_f}^\infty dx \frac{\langle \sigma v_{\rm rel}\rangle}{x^2})$ is given by~\cite{Nomura:2017jxb, Nomura:2017wxf}
\begin{align}
J(x_f)&=\int_{x_f}^\infty dx\left[ \frac{\int_{4M_X^2}^\infty ds\sqrt{s-4 M_X^2} [W_{Z'}(s)+W_{z_{\varphi'}}(s)] K_1\left(\frac{\sqrt{s}}{M_X} x\right)}{16  M_X^5 x [K_2(x)]^2}\right],\\ 
W_{Z'}(s)
\approx &\frac{4(s-4M_X^2)}{3\pi}  \left| \frac{5g_{BL}^2}{s-m_{Z'}^2+i m_{Z'} \Gamma_{Z'}}\right|^2 
\sum_f  \sqrt{1-\frac{4 m_{f}^2}{s}} (s+2 m^2_{f})|Q_{BL}^{f}|^2,\label{eq:cs-zp}\\
 W_{\alpha_{G}}(s)
\simeq &
\frac{|M_{X}|^4}{64\pi v^4_{\varphi_{10}}} 
\left[
(3s^2-4M_X^4) \left( \frac{\pi}{2sM_X^2}\sqrt{\frac{M_X^4}{4sM_X^2-s^2}} - \frac{\tan^{-1}\left[\frac{s-2M_X^2}{\sqrt{s(4M_X^2-s)}}\right]}{s^{3/2}\sqrt{4M_X^2-s}} \right)-4\right],  \label{eq:cs-gb}
\end{align}
where we assumed $Z'$ boson and scalar bosons are heavier than $X$ to forbid corresponding annihilation processes kinematically, for simplicity.
Here decay width of $Z'$ is given by Eq.~(\ref{eq:widths}) where $Z'$ can decay into $2X$, if kinematically allowed.
%
We find that two characterized solutions of measured relic density $\Omega h^2\approx0.12$~\cite{Ade:2013zuv} in the above formula.
The first one is a sharp region at around $M_X\sim m_{Z'}/2$, that is a resonant solution from the contribution $2X\to Z'\to f\bar f$ in Eq.~(\ref{eq:cs-zp}).
The second one is  the region in lighter mass of DM that mainly arises from the contribution $2X\to 2\alpha_G$ in~Eq.(\ref{eq:cs-gb}).
In the former case DM mass is around TeV scale to obtain right relic density due to the collider constraints for $Z'$ mass while
in the latter case DM mass can be $\mathcal O(10)$ GeV to $\mathcal O(100)$ GeV which depend on the coupling factor $M_X/v_{\varphi_{10}}$;
for more details, see, {\it e.g.}, Refs.~\cite{Nomura:2017jxb, Nomura:2017wxf}.


\section{Conclusion}
\label{sec:conclusion}

We have systematically explored the origins of neutrino textures in the canonical seesaw model with two right-handed neutrinos 
based on global U(1)$_{\mu-\tau}$ flavor symmetry, and smaller non-Abelian flavor symmetries,
and we have shown several promising  symmetries to find predictive textures, $U(1)_{\mu-\tau}$ and $D_4$, depending on appropriate charge assignments of our fields.
{Moreover, we have found that $D_4$ symmetry can realize a predictive texture $b$ only.}
Then we have proposed a concrete model based on local $U(1)_{B-L}$ and $D_4$ symmetries that involves a dark matter candidate and extra gauge boson. 
To show properties of the model, we have analyzed the neutrino physics, collider physics regarding $Z'$ boson and relic density of dark matter.
We have shown that constraints for $Z'$ mass and interactions can be relaxed when exotic scalar modes of $Z'$ decay are kinematically open, and 
relic density of dark matter can be explained by annihilation mode via $Z'$ exchange and/or annihilation into physical Goldstone bosons.

\section*{Acknowledgments}
\vspace{0.5cm}
{\it T.~K. was is supported in part by MEXT KAKENHI Grant Number JP17H05395.\\
This research is supported by the Ministry of Science, ICT and Future Planning, Gyeongsangbuk-do and Pohang City (H.O.). 
H. O. is sincerely grateful for the KIAS member, too.}

\section*{Appendix}
Here we show the valid multiplication rules for $D_4$ group that consists of four irreducible singlets $1,1',1'',1'''$ and one
irreducible doublet $2$, where we have used a real representation~\cite{Vien:2013zra};
\begin{align}
&\left[\begin{array}{c}
x_1 \\ 
 x_2 \\ 
\end{array}\right]_2
\otimes 
\left[\begin{array}{c}
y_1 \\ 
 y_2 \\ 
\end{array}\right]_2=
(x_1y_1+x_2y_2)_1\oplus (x_1y_1-x_2y_2)_{1'}\oplus (x_1y_2+x_2y_1)_{1''}\oplus (x_1y_2-x_2y_1)_{1'''}.
\end{align}
The other relations are given by $2\otimes 1(1',1'',1''')=2$, $1'\otimes 1'(1'',1''')=1(1''',1'')$, $1''\otimes 1''(1''')=1(1')$, and , $1'''\otimes 1'''=1$ in ref.~\cite{Ishimori:2010au}.

\end{document}